\documentstyle[aps,prl,multicol,amsfonts,psfig]{revtex}

\newcommand{\beq}{\begin{equation}}
\newcommand{\eeq}{\end{equation}}
\newcommand{\bea}{\begin{eqnarray}}
\newcommand{\eea}{\end{eqnarray}}
\newcommand{\bml}{\begin{mathletters}}
\newcommand{\eml}{\end{mathletters}}

\input epsf

\begin{document}

\draft 

\title{Phase transitions as topology changes in
configuration space: an exact result}

\author{Lapo Casetti$^{1,}$\cite{lapo}, E.~G.~D.~Cohen$^{2,}$\cite{egd} and
Marco Pettini$^{1,3,}$\cite{marco}}

\address{$^1$Istituto Nazionale per la Fisica della Materia (INFM), UdR Firenze,
Largo Enrico Fermi 2, I-50125 Firenze, Italy} 

\address{$^2$The Rockefeller University, 1230 York Avenue, New York, New York 
10021 - 6399} 

\address{$^3$Osservatorio Astrofisico di Arcetri, Largo Enrico Fermi 5,
I-50125 Firenze, Italy} 

\date{April 14, 2001}

\maketitle

\begin{abstract} 
The phase transition in the mean-field $XY$ model
is shown analytically to be related to a topological change in its configuration
space. Such a topology change is completely described by means of Morse theory
allowing a computation of the Euler characteristic---of suitable submanifolds
of configuration space---which shows a sharp discontinuity at the phase
transition point, also at finite $N$.  The present analytic result provides,
with previous work, a new key to a possible connection of topological changes
in configuration space as the origin of phase transitions in a
variety of systems. 
\end{abstract} 


Phase transitions (PTs) are one the most striking phenomena in nature.
They involve sudden qualitative physical changes,  
accompanied by sudden changes in the thermodynamic quantities  
measured in experiments. From a mathematical point of view, both 
qualitative and quantitative changes at PTs are conventionally described by 
the loss of analyticity of the probability measures and of the 
thermodynamic functions.
According to statistical mechanics, in the grandcanonical and canonical
ensembles, such a nonanalytic behavior can exist
only in the thermodynamic limit, i.e., in the rather
idealized case of a system with $N \to \infty$ degrees of freedom
\cite{Kramers}. PTs in real systems would then be the ``shadow'', at finite but
large $N$, of this idealized behavior. The way nonanalytic behavior can emerge
as $N \to \infty$ due to singularities of the probability measures of the
statistical ensembles has been rigorously studied by Yang and Lee in the 
grandcanonical ensemble \cite{YL} and by Ruelle, Sinai and others \cite{RS} in
the canonical ensemble; however, to the best of our knowledge, no equally
rigorous approaches to this problem exist in the microcanonical ensemble
\cite{micro}. Moreover, the necessity of taking the $N\to\infty$ limit to speak
of PTs seems less satisfactory today, since there is growing experimental
evidence of PT phenomena in systems with {\em small} $N$ (e.g.,
atomic clusters and nuclei, polymers and proteins, nano and mesoscopic
systems). 
There is also a deeper reason why the conventional approach to PTs is not yet 
completely satisfactory. Consider a classical system described by 
a Hamiltonian ${\cal H} = K(\pi) + V(\varphi)$  where $K(\pi) =
\frac{1}{2}\sum_{i=1}^N \pi_i^2$ is the kinetic energy, $V(\varphi)$ 
is the potential
energy and $\varphi \equiv \{\varphi_i \}$ and $\pi \equiv \{\pi_i \}$'s are, 
respectively, the canonical
conjugate  coordinates and momenta. Although in principle all the information
on the statistical properties is contained in the function $V(\varphi)$, 
no general 
result is available to specify which features of $V(\varphi)$ are necessary and
sufficient to entail the existence of a PT. 
This is the more surprising since in many
cases, knowing {\em a priori} that a system undergoes a PT, several relevant
properties of the PT can be predicted just in terms of very general features 
of $V(\varphi)$ (e.g., by means of renormalization-group techniques).
In the light of some
recent results \cite{cccp,franzosi,xymf,phi4,CSCP,theorem,physrep} an alternative
new approach approach seems actually possible by 
resorting to {\em topological} concepts: PTs would then be related
to {\em topology changes} (TCs) of suitable submanifolds of configuration
space, defined by the potential energy $V(\varphi)$. The connection between
topology of configuration space and PTs can be heuristically understood as 
follows. The microcanonical volume $\Omega(E)$
can be written, at total energy $E$, as a convolution integral \cite{Ruelle}
\begin{equation}
\Omega (E)= \int_0^E dK\, \text{Vol} \left({\Bbb S}_{(2K)^{1/2}}^{N-1} \right) 
\text{Vol} \left(M_{E-K} \right) ~,
\end{equation}
where ${\Bbb S}_{R}^{n}$ is an $n$-dimensional hypersphere of radius $R$, $K$
represents the kinetic energy  
and $M_V = M_{E-K}$ 
is the subset of the configuration space $M$ contained within
the equipotential hypersurface of level $V=E-K$. As $K$ is varied, 
all the hyperspheres are trivially topologically equivalent. On the contrary,
the topology of $M_{E-K}$ can change for some value of $K$, also at
finite (and even small) $N$. 
The larger $N$, the better
$\Omega(E)$ is approximated by the product of volumes 
$\text{Vol}\left({\Bbb S}_{\langle 2K \rangle^{1/2}}^{N-1}\right) \, 
\text{Vol}\left(M_{\langle V \rangle}\right)$, where 
$\langle K \rangle$ and $\langle V \rangle$ are, respectively,
the kinetic and potential energy averaged over the microcanonical equilibrium
distribution
and $\langle K \rangle = E - \langle V \rangle$. 
An abrupt TC of the $M_{E-K}$ can yield singular derivatives of
$\Omega(E)$ \cite{theorem,CSCP1}; if it is persistent with $N$, such a TC 
will 
result in a loss of analyticity of the thermodynamic observables, only in the
$N\rightarrow\infty$ limit. 
However, in the present approach the loss of 
analyticity is due to a deeper primitive topological cause of a PT, 
which is already present at finite $N$ \cite{note_can}.

Although it has been proven that TCs in configuration space are a {\em
necessary} condition for PTs to occur \cite{theorem}, it is known that {\em not
all} TCs are related to PTs  \cite{physrep}, and no general argument is yet at
hand to define the {\em sufficient} conditions under which a TC is actually
related to a PT. The study of particular models is then crucial to obtain clues
towards the general solution of this problem. The main purpose of this Letter
is to give a complete analytical characterization of the TCs in the
configuration space of the so-called mean-field $XY$ model, which may indicate
the difference between the TC  related to the PT and other TCs.  Before we
discuss this model, we first summarize a few needed facts about topology.  

The TCs we are referring to are those transformations which map a manifold onto
a new one which is not diffeomorphic to the previous one, i.e., which cannot be
mapped back to it by means of a differentiable transformation. A TC
is therefore any transformation which ``breaks the fabric'' of a
manifold: making a hole---without boundary---in a sphere  tranforms it into a
torus, and there is no smooth way to transform a torus back to a sphere. Morse
theory \cite{morse} provides a way of classifying TCs of manifolds, and links
{\em global} topological properties with {\em local}  analytical  properties of
smooth functions defined on them, so that they can be used as a  practical tool to
study their topology. Given a (compact) $N$-dimensional  manifold $M$ and a
smooth function $f: M \mapsto {\Bbb R}$, a point  $\overline{x} \in M$ is called a {\em
critical point} of $f$ if $df = 0$, while the  value $f(\overline{x})$ is called a {\em
critical value}. A level set  $f^{-1}(a) = \{ x \in M : f(x) = a \}$ of $f$ is
called a {\em critical level} if $a$ is a critical value of $f$, i.e., if there
is at least one  critical point $\overline{x} \in f^{-1}(a)$. The function $f$ is called
a {\em Morse function} on $M$ if its critical points are all nondegenerate, i.e.,
if the Hessian of $f$ at $\overline{x}$ has only nonzero eigenvalues, so that the critical
points $\overline{x}$ are isolated.  We now consider the configuration space of 
a classical system as our manifold $M$, and the potential energy per particle 
${\cal V}(\varphi) = V(\varphi)/N$
as our Morse function. Then the submanifolds $M_v$ of $M$ whose
topology we want to investigate are 
\beq 
M_v = {\cal V}^{-1} (-\infty,v] =  \{ \varphi \in M : {\cal V}(\varphi) 
\leq v\}~, 
\label{submanifolds}
\eeq
i.e., the same as the $M_V = M_{E-K}$ defined above 
(where in $M_v$, $V$ has been rescaled by $\frac{1}{N}$, i.e., $v = V/N$,
in order to make the comparison of systems with different $N$ easier). 
All the submanifolds $M_v$  of $M$, with increasing $v$,  
have the same topology until a
critical level ${\cal V}^{-1}(\overline{v})$ is crossed. Here the
topology of $M_v$ changes in a way 
completely determined by the local properties of the Morse function: at any
critical level a $k$-handle $H^{(k)}$ is attached \cite{handles}, where $k$ is
the {\em index} of the critical point, i.e., the number of negative eigenvalues
of the Hessian matrix of $\cal V$. Notice that if there are $m>1$ critical
points on the same critical level, with indices $k_1,\ldots,k_m$, then the
TC is made by attaching $m$ disjoint handles
$H^{(k_1)},\ldots,H^{(k_m)}$.
This way, by increasing $v$, the full configuration space $M$ can be 
constructed sequentially from the $M_v$.
Knowing the index of all the critical points below a given level $v$, we can 
obtain {\it exactly} the Euler characteristic of the manifolds $M_v$, given by
\beq
\chi (M_v) = \sum_{k = 0}^N (-1)^k \mu_k(M_v)~,
\label{chi_morse}
\eeq
where the {\em Morse number} $\mu_k$ is the number of critical points of $\cal V$
which have index $k$ \cite{note_chi}. The Euler characteristic $\chi$ is a
{\em topological invariant}: any change in $\chi(M_v)$ implies a
TC in the $M_v$.

Thus, in order to detect and characterize topological 
changes in $M_v$ we have to find the 
critical points and the critical 
values of ${\cal V}$, which means solving the equations
\beq
\frac{\partial {\cal V}(\varphi)}{\partial \varphi_i} = 0~, 
\qquad i = 1,\ldots,N~,
\label{crit_eqs}
\eeq
and to compute the indices of {\em all} the critical points of $\cal V$,
i.e., the number of negative eigenvalues of its Hessian
\beq
H_{ij} = \frac{\partial^2 {\cal V}}{\partial \varphi_i \partial \varphi_j}  
\qquad i,j = 1,\ldots, N\,.
\label{hess}
\eeq 
In the case of the mean-field $XY$ model, which  describes a 
system of $N$ equally coupled
planar classical rotators \cite{Antoni}, 
due to the mean-field 
character of the interactions, such a calculation can be done
in a completely analytical way. This allows then a
new discussion of the relationship 
between TCs and the PT of this model, whose potential energy is 
\beq
V(\varphi) = \frac{J}{2N}\sum_{i,j=1}^N 
\left[ 1 - \cos(\varphi_i - \varphi_j)\right] -h\sum_{i=1}^N \cos\varphi_i ~,
\label{V}
\eeq
where $\varphi_i \in [0,2\pi]$ is the rotation angle of the $i$-th rotator 
and $h$ is an external field. The model 
describes also a planar ($XY$) Heisenberg system with interactions of 
equal strength among all the classical spins
${\bf s}_i = (\cos\varphi_i,\sin\varphi_i)$. 
We consider only the ferromagnetic case $J >0$; 
for the sake of simplicity, we set $J=1$. 
In the limit $h \to 0$, the system has a continuous 
PT, with classical critical exponents, at $T_c = 1/2$, or 
$\varepsilon_c = 3/4$, 
where $\varepsilon = E/N$ is the total energy per particle \cite{Antoni}. 
Defining the magnetization vector per particle
${\bf m} = \left(m_x,m_y\right)$, where
$m_x = \frac{1}{N}\sum_{i=1}^N \cos\varphi_i$, 
$m_y = \frac{1}{N}\sum_{i=1}^N \sin\varphi_i$,
the potential energy $V$ can be written as a function of $\bf m$ as:
\beq
V(\varphi) = V(m_x,m_y) = \frac{N}{2} (1 - m_x^2 - m_y^2) 
- hN\, m_x~.
\label{V(m)}
\eeq
The range of values of the potential energy per particle,
${\cal V} = \frac{V}{N}$, is then 
$-h \leq {\cal V} \leq \frac{1}{2} + \frac{h^2}{2}$.

The configuration space $M$ of the model is an $N$-dimensional torus, being
parametrized by the $N$ angles $\{\varphi_i \} = \varphi_1,\dots ,\varphi_N$. 
We now study the topology of the family of submanifolds $M_v$ 
for this model. First, since  TCs of $M_v$ can occur only 
at  critical points of $\cal V$, there are
no TCs when $v > \frac{1}{2} + \frac{h^2}{2}$, 
i.e., all the $M_v$'s with $v > \frac{1}{2} + \frac{h^2}{2}$ 
must be diffeomorphic to the whole $M$, that is, they must be $N$-tori.
Then one has to find all the solutions of
Eqs.~(\ref{crit_eqs}), which can be rewritten in the form \cite{xymf,physrep}
\beq
(m_x + h) \sin\varphi_i - m_y \cos\varphi_i = 0 ~ , 
\qquad i = 1,\ldots,N~.
\label{crit_eqs_i}
\eeq
As long as $(m_x + h) \not = 0$ and $m_y \not = 0$ ($m_x$ and $m_y$ are both
zero only on the level $v = 1/2 + h^2/2$), the solutions of
Eqs.\ (\ref{crit_eqs_i}) are all those configurations for which the angles 
$\varphi_i$ are
either $0$ or $\pi$. These configurations
correspond to a value of $v$ which depends only on the number of angles $n_\pi$
which are equal to $\pi$, and using Eq.\ (\ref{V(m)}) one obtains 
\beq
v(n_\pi) = \frac{1}{2}\left[1 - \frac{1}{N^2}\left(N - 2n_\pi\right)^2\right] - 
\frac{h}{N}\left(N - 2n_\pi\right)\,, 
\label{v(n)}
\eeq
where $0 \leq n_\pi \leq N$. 
We have thus shown that as $v$ changes from its minimum $-h$ 
(corresponding to $n_\pi = 0$) to  
$\frac{1}{2}$ (corresponding to $n_\pi = \frac{N}{2}$)
the manifolds $M_v$ undergo a sequence of topology
changes at the $N$ critical values $v(n_\pi)$ given by Eq.~(\ref{v(n)}). 
There might be a TC
also at the last (maximum) critical value  
$v_c = \frac{1}{2} +
\frac{h^2}{2}$. However, the above argument does not prove it, 
since the critical points of $\cal V$ corresponding to this critical 
level may be degenerate \cite{note_degeneracy}, 
so that on this level $\cal V$ would not be a proper Morse
function. Then a critical value $v_c$ is still a necessary condition for the
existence of a TC, but it is no longer sufficient.  
However, as argued in Refs.~\cite{xymf,physrep}, 
it is just this TC at $v_c$ which should be
related to the thermodynamic PT of the mean-field XY model. 
For, the temperature $T$, the 
energy per particle $\varepsilon$ and the average potential energy per 
particle $u = \langle {\cal V} \rangle$ obey, 
in the thermodynamic limit, the equation $2\varepsilon = T + 2u(T)$,
where we have set Boltzmann's constant equal to 1. Substituting in this equation
the values of the critical energy per particle  and of the critical
temperature we get  $u_c = u(T_c) = 1/2$; as $h \to 0$, $v_c \to \frac{1}{2}$, 
so that $v_c = u_c~$.
Thus a TC in $M$ occurring at this $v_c$, 
where $v_c$ is {\em independent} of $N$, is connected with the
PT in the limit $N\to\infty$, and $h \to 0$,
when indeed thermodynamic PTs are usually defined. 

\begin{figure}
\epsfysize= 6 truecm 
\begin{center}
\mbox{\epsfbox{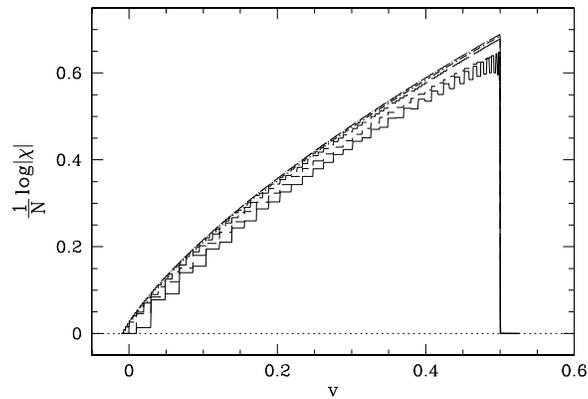}} 
\end{center}
\caption{Plot of $\log(|\chi|(M_v))/N$ as a function of $v$ for $h = 0.01$ and 
increasing $N =$ 50,100,200,400,800 (from bottom to top).}
\label{fig_chi}
\end{figure}

Let us now prove that a TC at $v_c$ actually exists and try to understand why it
is different from the other TCs, i.e., those occurring at $0 \leq v < v_c$. 
To do that we characterize all the TCs
occurring at the critical values $0 \leq v < v_c$ using Morse theory, 
computing the {\em indices} of the critical 
points of $\cal V$. 
At these points, where the angles are either $0$ or $\pi$, we can write the
Hessian matrix (\ref{hess}) in the form
\beq
NH = D + B
\label{matrixH}
\eeq
where $D$ is diagonal, $D = {\rm diag}(\delta_i)$, with
\beq
\delta_i = (m_x + h) \cos\varphi_i\,,
\eeq
and the elements of $B$, $b_{ij}$, can be written in terms of a vector
$\sigma$ whose $N$ elements are either $1$ or $-1$:
\beq
b_{ij} = - \frac{1}{N} \sigma_i \sigma_j\,,
\label{bij}
\eeq
where $\sigma_i = +1$ (resp.~$-1$) if $\varphi_i = 0$ (resp.~$\pi$). The matrix
$B$ has only one nonzero eigenvalue. This implies that the number of negative
eigenvalues of $H$ equals the number of negative eigenvalues of $D$ $\pm 1$ 
\cite{note_proof}, 
so that as $N$ gets large we can conveniently approximate the index of the
critical point with the number of negative $\delta$'s at $x$. 
At a given critical point, with given $n_\pi$,  
the eigenvalues of $D$ are 
\bml
\bea
\delta_i & = & m_x + h \qquad \qquad i  =  1,\ldots,N - n_\pi\,; \\
\delta_i & = & -(m_x + h) \qquad ~ i  =  N - n_\pi + 1,\ldots,N\,,
\eea
\eml
where the $x$-component of the magnetization vector is 
$m_x = 1 - \frac{2n_\pi}{N}$, 
so that $m_x >  0$ (resp.~$<0$) if $n_\pi \leq \frac{N}{2}$ (resp.~$ >
\frac{N}{2}$). Then, if the external field $h$ is sufficiently small, and 
denoting by ${\rm index}(n_\pi)$ the index of a critical point with given 
$n_\pi$ we can write
\bml
\bea
{\rm index}(n_\pi) & = & n_\pi \qquad \qquad  {\rm if}~ n_\pi \leq
\hbox{$\frac{N}{2}$}~, \\
{\rm index}(n_\pi) & = & N - n_\pi \qquad  {\rm if}~ n_\pi > \hbox{$\frac{N}{2}$}~.
\eea
\label{index(n)}
\eml
The number $C(n_\pi)$ of critical points having a 
given $n_\pi$, which is the number of distinct strings of $0$'s and $\pi$'s  
of length $N$ having $n_\pi$ occurrences of $\pi$, is given by
the binomial coefficient $C(n_\pi) = {N \choose n_\pi}$. 
Thus, at any critical level $-h \leq v(n_\pi) \leq \frac{1}{2}$, where
$v(n_\pi)$ is given by Eq.~(\ref{v(n)}),
a topological 
change in $M_v$ occurs, which is made up of attaching $C(n_\pi)$ $k$-handles,
where $k(n_\pi) = {\rm index}(n_\pi)$ given in Eq.~(\ref{index(n)}). Here
$n_\pi$ as a function of $v$ can be obtained by solving Eq.~(\ref{v(n)}), 
yielding
\beq
n_\pi^{(\pm)}(v) = \text{int}\, \left\{ \hbox{$\frac{N}{2}$}\left[ 1 + h \pm 
\sqrt{h^2 - 2 \left(v - \hbox{$\frac{1}{2}$} \right)} \right] \right\} \,,
\label{n(v)}
\eeq
where $\text{int}\,\{ a \}$ stands for the integer part of $a$.
Eqs.~(\ref{index(n)}) 
and (\ref{n(v)}) allow us to write the Morse
numbers $\mu_k$ of the manifolds $M_v$, for $-h \leq v < \frac{1}{2} +
\frac{h^2}{2}$, as
\bea
\mu_k (v) & = &
\left[1 - \Theta(k - n^{(-)}_\pi(v))  
 +  \Theta(N - k - n^{(+)}_\pi(v)) \right] \nonumber \\
 & \times & {N \choose k} \, , \qquad k = 0,1,\ldots,N\,,
\label{mu_k}
\eea
where $\Theta(x)$ is the Heaviside theta function. 
We note that since $0 \leq n_\pi^{(-)} \leq \frac{N}{2}$ and 
$ \frac{N}{2} + 1 \leq n_\pi^{(+)} \leq N$,  Eq.\ (\ref{mu_k}) implies
\beq
\mu_k (v) = 0 \qquad \forall\,k > \hbox{$\frac{N}{2}$}~,
\eeq
i.e., no critical points with index larger than $\frac{N}{2}$ exist as long as 
$v < \frac{1}{2} + \frac{h^2}{2}$. 
On the other hand, for $v > \frac{1}{2} + \frac{h^2}{2}$,  
$M_v$ must be 
an $N$-torus ${\Bbb T}^N$, and for {\em any} Morse function on such a manifold
one has \cite{torus_note} 
\beq
\mu_k ({\Bbb T}^N) \geq {N \choose k} \qquad k = 0,1,\ldots,N\,.
\eeq
Thus, as $\frac{1}{2} \leq v < \frac{1}{2} + \frac{h^2}{2}$ 
the manifold is only ``half'' an $N$-torus, and
since we know that for $v > \frac{1}{2} + \frac{h^2}{2}$, $M_v$ {\em is} an
$N$-torus, we conclude that at 
$v = v_c = \frac{1}{2} + \frac{h^2}{2}$ a TC {\em
must} occur, which involves the attaching of
$N\choose k$ $k$-handles for each $k$ ranging from $\frac{N}{2} + 1$ to $N$. 
This is
surely a ``big'' TC: all of a sudden, ``half'' an $N$-torus
becomes a full $N$-torus. 
Now we can use Eqs.\ (\ref{chi_morse}), (\ref{n(v)}) and (\ref{mu_k}) to 
compute the 
numerical values of the Euler characteristic of the manifolds $M_v$ as a 
function of $v$: 
it turns out that $\chi$ jumps from positive to negative
values, so that it is easier to look at $|\chi|$. 
In Figure \ref{fig_chi},  $\log(|\chi|(M_v))/N$ is plotted as a function of
$v$ for various values of $N$ ranging from 50 to 800. 
The ``big'' TC occurring at the maximum value $v_c$ 
of $\cal V$, which
corresponds in the thermodynamic limit to the PT, implies a
discontinuity of $|\chi|$, jumping from a big value (${\cal O}(e^N)$ 
in our case) to zero, which is the value of $\chi$ for an $N$-torus.

The analytical results we have presented provide a possible hint about  what
could be the {\em sufficient} conditions for a TC in configuration space  to
yield a PT.  In the model studied here, the $N$ TCs which are not related to
the PT  involve the simultaneous attachment of handles which are all of the
{\em same}  type, while that occurring at $v_c$ is the simultaneous attaching
of handles of ${\cal O}(N)$ {\em different} types. Hence we might {\em
conjecture} that this is  a sufficient condition for a TC to be in  one-to-one
correspondence with a thermodynamical PT, also in other models.  Further 
confirmation or refinements of this conjecture may provide the basis for a full
theory of the origin of PTs based on the topology of configuration space as
encoded in the potential energy. It is worth noticing that this might be useful
not only to study PTs in the microcanonical ensemble and/or in finite systems,
but also in systems which are difficult to study with conventional approaches,
like disordered ones or glasses. For glass-forming liquids topological concepts
have been recently invoked \cite{Angelani}.

We warmly thank G.~Vezzosi for discussions and suggestions. 
This work is part of the INFM PAIS {\em Equilibrium and Non-Equilibrium 
Dynamics of Condensed Matter Systems}. EGDC is indebted to 
the Office of Basic Energy Sciences 
of the US DOE under contract DE-FG02-88-ER13847 and to the IS Sabbatical 
of the INFM-Sez.~G.

\begin{references}
\bibitem{CSCP1} In Ref.~\protect\cite{CSCP} the connection between 
derivatives of $\Omega$ and topological invariants of the equipotential 
hypersurfaces (i.e., the boundaries of the $M_{E - K}$) is discussed.
\bibitem{note_can} This argument can be extended to the canonical and
grandcanonical ensembles 
since the partition and grand-partition functions are Laplace transforms of  
$\Omega (E)$.
\bibitem{handles} Morse theory builds up the submanifolds by 
``attaching handles'' \protect\cite{morse}.
Handles are products of disks: a $k$-handle $H^{(k)}$ in
$N$ dimensions ($0\leq k \leq N$) is the product of a $k$-dimensional disk 
-- or $k$-disk -- and an $(N-k)$-disk, i.e., defined by 
$H^{(k)} = D^{(k)} \times D^{(N-k)}$.
\bibitem{note_chi} This follows from the so-called 
Morse inequalities, stating that the Betti numbers (the fundamental topological
invariants of a manifold) are bounded by the Morse numbers, and from that the
Euler characteristic is the alternate sum of the Betti numbers 
(see e.g.~Ref.~\protect\cite{morse}).
\bibitem{note_degeneracy} The solutions of the two equations in $N$ variables
$m_x = m_y = 0$ need not to be isolated.
\bibitem{torus_note} The $\mu_k$'s are upper limits of the Betti numbers
(see \protect\cite{note_chi}); the $k$-th Betti number of an $N$-torus is 
$N\protect\choose k$.
\end{references}
\end{document}